\input phyzzx
\hsize=417pt 
\sequentialequations
\Pubnum={ EDO-EP-14}
\date={ \hfill July 1997}
\titlepage
\vskip 32pt
\title{ Quantum Evaporating Black Holes in Higher Dimensions}
\vskip 12pt
\author{Ichiro Oda \footnote\dag {E-mail address: 
ioda@edogawa-u.ac.jp}}
\vskip 12pt
\address{ Edogawa University,                                
          474 Komaki, Nagareyama City,                        
          Chiba 270-01, JAPAN     }                          
%
%
%
%
%
\abstract{ We study an $N + 1$ dimensional generalization of the 
Schwarzschild black hole from the quantum mechanical viewpoint. It is 
shown that the mass loss rate of this higher dimensional black hole due 
to the black hole radiation is proportional to $1 \over r^{N-1}$ where 
$r$ is the radial coordinate. This fact implies that except in four 
dimensions the quantum formalism developed in this letter gives a 
different result with respect to the mass loss rate from the 
semiclassical formalism where the Stefan-Bolzmann formula for a blackbody 
radiation is used. As shown previously in the study of the four 
dimensional Schwarzschild metric, the wave function in this case has 
also quite curious features that it is singular owing to strong quantum 
fluctuation of the gravitational field at the singularity while 
completely regular in the other regions of the spacetime.
}
\endpage
%
%
%

\def\sp(#1){\noalign{\vskip #1pt}}

%
%
%
%
%
\topskip 30pt
\par

Despite much impressive effort, establishing a quantum field theory which 
unifies the gravitational interaction with the other interactions 
existing in nature remains to be one of unsolved problems in the modern 
theoretical physics. Motivated by the development of superstring theories 
[1], it seems that most recent works in this search have been directed at 
studying theories where the number of spacetime dimension is greater than 
four. In particular, in more recent works it is widely expected that 
black holes would play a very important role in understanding the 
non-perturbative features of quantum gravity [2]. Hence examining quantum 
aspects of black holes in $N + 1$ dimensions with $N \geq 3$ is certainly 
of importance in obtaining useful informations about quantum gravity when 
we attempt to construct a unified theory in future.

In this letter, we would like to study the Hawking radiation [3] of the 
higher dimensional analog of the $3 + 1$ dimensional Schwarzschild black 
hole [4, 5] in terms of the recently developed formalism [6-11] which 
goes beyond the semiclassical analysis [3] and is purely quantum 
mechanical at least in the geometry involving a black hole. Although we 
deal with a specific higher dimensional black hole in general relativity 
for the sake of simplicity, it is straightforward to apply the present 
formalism to the more complicated black holes with several nontrivial 
charges which have recently been found in the low energy effective theory 
of superstrings [2].

Our motivations in this letter are twofold. On the one hand, by extending 
the previous formalism constructed in dimensions equal to or lower than 
four to the higher dimensional Schwarzschild black hole  we would like to 
understand some quantum aspects of the black hole evaporation in an 
arbitrary dimension. On the other hand, it was shown in the previous 
works [6-11] that in two and four dimensions the mass loss rate of quantum 
evaporating black holes coincides with the result obtained in the 
semiclassical analysis. Then one is naturally led to ask whether this 
coincidence remains true even in higher spacetime dimensions. We will see 
that this is not the case in general. This fact makes it clear that our 
quantum mechanical approach is surely different from the semiclassical one.

The classical action which we consider has the form   
$$ \eqalign{ \sp(2.0)
S = \int  d^{N+1} x \sqrt{-^{(N+1)}g} \ \bigl( \ {1 \over 16 \pi G} 
{}^{(N+1)}R - {1 \over  8 \pi} {}^{(N+1)}g^{\mu\nu} \partial_{\mu} \Phi 
\partial_{\nu} \Phi \  \bigr),
\cr
\sp(3.0)} \eqno(1)$$
where $(N + 1)$ is put on the metric tensor and the curvature scalar to 
distinguish the $N + 1$ dimensional quantities from the two dimensional 
ones appearing in what follows. We follow the conventions adopted 
in the MTW textbook [12] and use the natural units $G = \hbar = c = 1$. 
The Greek indices $\mu, \nu, ...$ take the values 0, 1, 2, ... N, while 
the Latin indices $a, b, ...$ run over the two dimensional values 0 and 1. 

Adopting a general spherically symmetric ansatz [13]
$$ \eqalign{ \sp(2.0)
ds^2 &= {}^{(N+1)}g_{\mu\nu} dx^{\mu} dx^{\nu},
\cr
     &= g_{ab}(x^c) dx^a dx^b + \phi^2 (x^c)  d\Omega_{N-1}^2,  
\cr
\sp(3.0)} \eqno(2)$$
with 
$$ \eqalign{ \sp(2.0)
d\Omega_{N-1}^2 = d\theta_2^2 + \sin^2\theta_2 d\theta_3^2 + \cdots + 
\prod_{i=2}^{N-1} \sin^2\theta_i d\theta_N^2,
\cr
\sp(3.0)} \eqno(3)$$
then (1) can be cast to be
$$ \eqalign { \sp(2.0)
S &= {\omega_{N-1} \over 16 \pi} \int  d^2 x \sqrt{-g} \ \bigl[  \phi^{N-1}  R 
+ ( N - 1) ( N - 2 ) \phi^{N-3} ( 1 + g^{ab} \partial_a \phi 
\partial_b \phi ) \bigr]
\cr
&\qquad- {\omega_{N-1} \over 8 \pi} \int  d^2 x \sqrt{-g} \ \phi^{N-1} g^{ab} 
\partial_a  \Phi \partial_b \Phi,
\cr
\sp(3.0)} \eqno(4)$$
where we have also set $\partial_{\theta_i} \Phi = 0$. In deriving (4), 
the following identities have been used:
$$ \eqalign { \sp(2.0)
{}^{(N+1)}R = R - 2 (N - 1) {1 \over \phi} g^{ab} \nabla_a \nabla_b 
\phi + (N - 1) (N - 2) {1 \over \phi^2} ( 1 - g^{ab} \nabla_a \phi 
\nabla_b \phi ),
\cr
\sp(3.0)} \eqno(5)$$
and 
$$ \eqalign { \sp(2.0)
{}^{(N+1)}g^{\theta_2 \theta_2} {}^{(N+1)}R_{\theta_2 \theta_2} = \cdots = 
{}^{(N+1)}g^{\theta_N \theta_N} {}^{(N+1)}R_{\theta_N \theta_N},
\cr
\sp(3.0)} \eqno(6)$$
which are features from spherical symmetry. And $\omega_{N-1}$ denotes the 
area of a unit $N-1$ sphere, which is defined by ${{2 \pi^{N \over 2}} 
\over \Gamma({N \over 2})}$. In this article, we confine ourselves to the 
cases of $N \geq 3$. $N = 1$ and $N = 2$ cases have been separately 
examined in our previous works [9]. From now on, as the classical action 
we shall make use of $\bar S = {4 \pi \over \omega_{N-1}} S$ in order to 
keep the correspondence with the $N = 3$ case [10]. 

Following the method in ref.[13], we are ready to construct the canonical 
formalism. Particularly, we can readily evaluate the Hamiltonian which 
turns out to be a linear combination of the Hamiltonian constraint $H_0$ 
and the momentum constraint $H_1$ as follows:
$$ \eqalign{ \sp(2.0)
H = \int dx^1 \ ( \alpha H_0 + \beta H_1 ), 
\cr
\sp(3.0)} \eqno(7)$$
with
$$ \eqalign{ \sp(2.0)
H_0 &= {1 \over 2 \sqrt{\gamma} \phi^{N-1}} p_{\Phi}^2 - {1 \over 4} (N 
- 1) (N - 2) \sqrt{\gamma} \phi^{N-3} \bigl[ 1 + {1 \over \gamma} ( 
\phi^\prime )^2 \bigr] + \partial_1 ( {\partial_1 (\phi^{N-1}) \over 2 
\sqrt{\gamma}} ) 
\cr
&\qquad+ {\phi^{N-1} \over 2 \sqrt{\gamma}} ( \Phi^\prime )^2 - {4 \over 
N-1} {\sqrt{\gamma} \over \phi^{N-2}} p_\phi p_\gamma + {4 (N - 2) \over 
N - 1} { \gamma \sqrt{\gamma} \over \phi^{N-1}} p_\gamma ^2, 
\cr
\sp(3.0)} \eqno(8)$$
$$ \eqalign{ \sp(2.0)
H_1 = {1 \over \gamma} \ p_\Phi \Phi^\prime + {1 \over \gamma} p_\phi 
\phi^\prime - 2  p_\gamma ^\prime - {1 \over \gamma} p_\gamma 
\gamma^\prime. 
\cr
\sp(3.0)} \eqno(9)$$
Here we have introduced the ADM parametrization [13]
$$ \eqalign{ \sp(2.0)
g_{ab} = \left(\matrix{ { - \alpha^2 + {\beta^2 \over \gamma}} & \beta \cr
              \beta & \gamma \cr} \right),
\cr
\sp(3.0)} \eqno(10)$$
and a prime denotes the differentiation with respect to $x^1$.

The $N+1$ dimensional generalization of the four dimensional 
Schwarzschild metric has been found by Tangherlini [4] (See ref.[5] for 
the detail of black holes in higher dimensional spacetimes) whose line 
element is
$$ \eqalign{ \sp(2.0)
ds^2 = - (1 - {C \over r^{N-2}}) dt^2 + {1 \over 1 - {C \over r^{N-2}}} 
dr^2 + r^2 d\Omega_{N-1}^2,  
\cr
\sp(3.0)} \eqno(11)$$
where the parameter $C$ is related to the black hole mass $M$ by
$$ \eqalign{ \sp(2.0)
C = {16 \pi G M \over (N - 1) \ \omega_{N-1}}.  
\cr
\sp(3.0)} \eqno(12)$$
Now for later convenience let us rewrite (11) into the form of the 
ingoing Vaidya metric [14]
$$ \eqalign{ \sp(2.0)
ds^2 = - (1 - {C \over r^{N-2}}) dv^2 + 2 dv dr + r^2 d\Omega_{N-1}^2,
\cr
\sp(3.0)} \eqno(13)$$
where $v$ is the advanced time coordinate and $C$ is the function of only 
$v$ coordinate. 

The canonical quantization of a system with the Vaidya metric (13) proceeds 
essentially as in the previous works [6-11]. Here, however, we would like 
to present the more detailed explanation than before. As a first step toward the 
canonical quantization, we have to select the  
following two dimensional coordinate 
$$ \eqalign{ \sp(2.0)
x^a = (x^0, x^1) = (v - r, \  r).
\cr
\sp(3.0)} \eqno(14)$$
This choice of the coordinate system is very crucial to reach the desired 
results in what follows. Next task is to fix the two dimensional 
reparametrization invariances by the gauge conditions 
$$ \eqalign{ \sp(2.0)
g_{ab} = \left(\matrix{ - (1 - {C \over r^{N-2}})   &  C \over 
              r^{N-2} \cr 
              C \over r^{N-2}  & 1 + {C \over r^{N-2}} } \right),
\cr
\sp(3.0)} \eqno(15)$$
where $C$, which is related to $M$ through (12), is a general function 
depending on the two dimensional coordinate $x^a$ so is $M$ also such a 
function. These gauge conditions are chosen such that the line element 
has almost the same form as the Vaidya metric. Finally, in order to make 
the metric (15) coincide with the ingoing Vaidya metric (13) precisely, 
let us make assumptions on the dynamical fields
$$ \eqalign{ \sp(2.0)
\Phi = \Phi(v), \ M = M(v), \ \phi =  r. 
\cr
\sp(3.0)} \eqno(16)$$
Here we wish to make a comment on (16). It is true that identifying the 
function $\phi$ with the mere coordinate $r$ is a great simplification 
leading to the minisuperspace model since it effectively kills the role 
as a dynamical field of $\phi$. However, this assumption can be seen as 
one method to gain a useful approximation to the exact Hamiltonian. 
Incidentally, consistency of (15) and (16) with the field equations 
stemming from (1) is checked by a lengthy but straightforward 
calculation.   

Now using various equations discussed above, we can find a remarkable 
equation 
$$ \eqalign{ \sp(2.0)
\sqrt{\gamma} H_0 &= \gamma  H_1,
\cr
             &= {1 \over \phi^{N-1}} \ p_{\Phi}^2  - \gamma \ p_\phi + 
             {(N - 1) (N - 2) \over 4} {C^2  \over r^{N-1}}, 
\cr
\sp(3.0)} \eqno(17)$$
where $\gamma = 1 + {C \over r^{N-2}}$ from (10) and (15). Note that (17) 
exactly reduces to the corresponding equation in the four dimensional 
Schwarzschild black hole when specified to $N = 3$ [10]. Replacing 
$p_{\Phi}$ and $p_{\phi}$ with $- i {\partial \over \partial \Phi}$ and 
$ - i {\partial \over \partial \phi}$, respectively, leads to the 
Wheeler-DeWitt equation.
$$ \eqalign{ \sp(2.0)
\bigl[ - {1 \over \phi^{N-1}} {\partial^2 \over \partial \Phi^2} + i \gamma  
{\partial \over \partial \phi} + {(N - 1) (N - 2) \over 4} {C^2  \over 
r^{N-1}} \bigr] \Psi = 0.  
\cr
\sp(3.0)} \eqno(18)$$
A special solution of this Wheeler-DeWitt equation can be found to be 
$$ \eqalign{ \sp(2.0)
\Psi = (  B_{+} e^{\sqrt{A} \Phi(v)} + B_{-} e^{-\sqrt{A} \Phi(v)} ) \ 
e^{ i {{ A 
- {1 \over 4} (N - 1) (N - 2) C^2} \over (N - 2) C} \log \gamma },
\cr
\sp(3.0)} \eqno(19)$$
where $A$ and $B_{\pm}$ are integration constants. This wave function has 
a perfectly similar behavior to that of the four dimensional 
Schwarzschild black hole [10]. Namely, at the spatial infinity $r  
\rightarrow  \infty$, $\gamma \rightarrow 1$ thus $\Psi$ consists of 
the ingoing matter field in the asymptotically flat spacetime while at 
the singularity $r \rightarrow  0$, $\gamma \rightarrow 
\infty$ so that $\Psi$ oscillates violently due to strong quantum effects 
associated with the gravitational degrees of freedom $\gamma$, which 
correspond classically to ``graviton''. In this respect, it is worthwhile 
to notice that $\Psi$ is completely regular over the whole region of 
spacetime except at the curvature singularity.

Then under a rather general definition of the expectation value, it is 
easy to evaluate the expectation value of the change rate 
of the black hole mass. The result is 
$$ \eqalign{ \sp(2.0)
< \partial_v M > = - {\omega_{N-1} \over 4 \pi} {A \over r^{N-1}}.
\cr
\sp(3.0)} \eqno(20)$$
This result as well as (18) and (19) also becomes equivalent to that of 
the four dimensional black hole when $N = 3$ [10].

So far we have described how to apply the formalism [6-11] to the higher 
dimensional Schwarzschild black hole [4, 5] and obtained a physically 
reasonable pictures with respect to both the wave function and the mass 
loss rate due to the Hawking radiation. Here a natural question arises as 
to what differences we would have when we compare the results obtained in 
the present formalism with the semiclassical results [3]. This is 
because both formalisms are certainly distinct: Our formalism is purely 
quantum mechanical in the sense that we have performed the canonical 
quantization of not only the gravitational field but also the matter 
field, while the semiclassical formalism [3] deals with the matter field 
as only the quantum field on the fixed background gravitational field. 
Therefore we expect both formalisms to provide us different behavior about 
the black hole evaporation. However, against this expectation,  
as far as the two and four dimensional black holes are concerned, we 
could not find any difference at least about the mass loss rate in the 
vicinity of the apparent horizon [6-11]. 
This appears to be somewhat strange. The purpose of this letter is to 
show that this is not always the case.

We begin by reviewing briefly the semiclassical formalism of the black 
hole radiation [15]. Since the radiation is of a blackbody nature [3], it 
is plausible to make use of the Stefan-Boltzmann law of blackbody 
radiation in evaluating an order estimate for thermal radiation from a 
black hole. Then it turns out that up to unimportant numerical constants, 
in $N + 1$ dimensions the mass loss rate of an evaporating black hole is 
given by 
$$ \eqalign{ \sp(2.0)
\partial_v M  &\sim - A_H T^{N+1} \sim - r_H^{N-1} T^{N+1} 
\cr
&\sim - (M^{1 \over N-2})^{N-1} (M^{- {1 \over N-2}})^{N+1} = - M^{- {2 
\over N-2}}, 
\cr
\sp(3.0)} \eqno(21)$$
where $A_H$ and $T$ respectively denote the area of the horizon and the 
Hawking temperature. Here we have assumed that particle creation occurs 
in the vicinity of the horizon. In addition, we have used the fact that 
the Hawking temperature $T$ and the horizon radius $r_H$ are proportional 
to $M^{- {1 \over N-2}}$ and $M^{1 \over N-2}$, respectively [4, 5]. 

On the other hand, our quantum mechanical formalism gave rise to the 
result (20) with respect to the mass loss rate. If we take $r$ to be the 
radius of the horizon, i.e., $r_H = C^{1 \over N-2}$, (20) yields 
$$ \eqalign{ \sp(2.0)
< \partial_v M >  \sim  - M^{- {{N - 1} \over N-2}}.
\cr
\sp(3.0)} \eqno(22)$$
Now by comparing (21) with (22), it is obvious that it is only in 3 + 1 
dimensions ($N = 3$) that the semiclassical and the quantum mechanical 
formalisms give the same dependency on the black hole mass up to 
numerical factors. This calculation clearly explains why both formalisms 
have provided the same mass loss rate in four dimensions in the previous 
works [6-11]. Incidentally, in two dimensions we have considered not 
the Einstein gravity but the dilaton gravity [16] so that the present 
analysis does not apply to this case in a direct manner [9].

In summary we have investigated some quantum aspects of a dynamical 
black hole corresponding to the Schwarzschild geometry in higher 
dimensions. Particularly, it was shown that there is the difference with 
respect to the mass loss rate between our quantum mechanical formalism 
and the conventional semiclassical formalism in $N + 1$ ($N \geq 4$) 
dimensions even if it is obscure in four dimensions. This observation in 
turn gives us a motivation that we should examine the present formalism 
further to understand various interesting properties associated with 
quantum black holes. One of the most attractive works in future seems to 
be to relax the assumption $\phi = r$ and construct the more general 
formalism than the present one, which would provide a more satisfactory 
treatment of the gravitational degrees of freedom. We wish to return to 
this problem in near future.    

\vskip 32pt
\leftline{\bf Acknowledgments}
\centerline{ } %
\par
We would like to thank K. Shiraishi for informations on ref.[4, 5]. This 
work was supported in part by Grant-in-Aid for Scientific Research 
from Ministry of Education, Science and Culture No.09740212.

\vskip 32pt
\leftline{\bf References}
\centerline{ } %
\par
\item{[1]} M.B.Green, J.H.Schwarz and E.Witten, Superstring Theory 
(Cambridge University Press, Cambridge, 1987). 

\item{[2]} G.T.Horowitz, gr-qc/9604051; J.M.Maldacena, hep-th/9607235.

\item{[3]} S.W.Hawking, Comm. Math. Phys. {\bf 43} (1975) 199.

\item{[4]} F.R.Tangherlini, Nuovo Cimento. {\bf 77} (1963) 636.

\item{[5]} R.C.Myers and M.J.Perry, Annals of Phys. {\bf 172} (1986) 304.

\item{[6]} A.Tomimatsu, Phys. Lett. {\bf B289} (1992) 283.

\item{[7]} M.Pollock, Int. J. Mod. Phys. {\bf D3} (1994) 579.

\item{[8]} A.Hosoya and I.Oda, Prog. Theor. Phys. {\bf 97} (1997) 233.

\item{[9]} I.Oda, gr-qc/9701058, 9703055, 9703056, 9704021, 9704084. 

\item{[10]} I.Oda, gr-qc/9705047. 

\item{[11]} P.Moniz, DAMTP R/97/23. 

\item{[12]} C.W.Misner, K.S.Thorne, and J.A.Wheeler, Gravitation (Freeman, 
1973).  

\item{[13]} P.Hajicek, Phys. Rev. {\bf D30} (1984) 1178 ; P.Thomi, B.Isaak  
and P.Hajicek, Phys.  Rev. {\bf D30} (1984) 1168.

\item{[14]} P.C.Vaidya, Proc. Indian Acad. Sci. {\bf A33} (1951) 264; 
P.Hajicek and W.Israel, Phys.  Lett. {\bf 80A} (1980) 9; W.A.Hiscock, 
Phys. Rev. {\bf D23} (1981) 2813, 2823.

\item{[15]} R.M.Wald, General Relativity (The University of Chicago Press, 
1984).

\item{[16]} C.G.Callan, S.B.Giddings, J.A.Harvey and A.Strominger, Phys. 
Rev. {\bf D45} (1992) R1005.

\endpage
%

%
\bye